\documentclass[11pt]{article}
\usepackage[T1]{fontenc}
\usepackage[utf8]{inputenc}
\usepackage{mathpazo}
\usepackage[margin=1in]{geometry}
\usepackage{amsmath,amssymb}
\usepackage{graphicx}
\usepackage{booktabs}
\usepackage[table]{xcolor}
\usepackage{caption}
\usepackage[super,sort&compress]{natbib}
\usepackage{authblk}
\usepackage{microtype}
\usepackage{array}
\usepackage{tabularx}
\usepackage{ragged2e}
\usepackage{placeins}
\usepackage[colorlinks=true,linkcolor=blue!50!black,citecolor=blue!50!black,urlcolor=blue!50!black]{hyperref}
\usepackage[capitalise,noabbrev]{cleveref}
\usepackage{subcaption}

\newcommand{\ABX}{ABX$_3$}

\title{Diffusion-Based Generative Screening of ABX$_3$ Perovskite Compositions for Bandgap-Targeted Materials Discovery}
\author[1]{Sunraj Patel}
\author[1,*]{Krishanu Biswas}
\affil[1]{Department of Materials Science and Engineering, Indian Institute of Technology Kanpur, India}
\affil[*]{Corresponding author: \href{mailto:kbiswas@iitk.ac.in}{\nolinkurl{kbiswas@iitk.ac.in}}}

\begin{document}
\maketitle

\begin{abstract}
  Crystal-generation models can explore compositions and atomic arrangements beyond a fixed set of crystal prototypes, but their usefulness for a specific materials family depends on whether the generated structures retain the defining topology and whether their predicted properties remain reliable across the sampled chemistry. Here, a pretrained diffusion model is used to generate candidate \ABX{} perovskites and to examine their structural, chemical and electronic-property distributions. A connectivity-based filter distinguishes three-dimensional corner-sharing perovskites from structures that share the same nominal ABX$_3$ stoichiometry. The accepted set contains substantial B-site multiplicity, including ordered double-perovskite configurations that are inaccessible to the two-prototype substitution reference used here. The generated set is enriched in chloride-, bromide- and iodide-containing compositions relative to the Materials Project, while all generated B-site elements are already represented in that database. On the downstream MP-trained ALIGNN scale, the predicted bandgap distribution is shifted towards 1.4 eV, but the target-window fraction (8.2\%) is close to the MP reference fraction (7.3\%); the separate MatterGen and ALIGNN label sources prevent this comparison from establishing target-condition enrichment. An ALIGNN model performs well on a formula-grouped random test split but does not outperform a constant baseline on a deliberately severe B-site-disjoint test. First-principles calculations on six generated structures show that relaxation can alter structural classification and that predicted and calculated gaps can differ appreciably. These results identify both the chemical-space coverage offered by crystal diffusion and the structural and predictive checks needed before its outputs can be treated as materials candidates.
\end{abstract}


\section{Introduction}

Perovskites are an important class of functional materials because their crystal chemistry accommodates a wide range of compositions and structural distortions. In the conventional \ABX{} structure, corner-sharing BX$_6$ octahedra form a three-dimensional framework, while the A-site cations occupy the larger spaces between the octahedra. Changes in composition, octahedral tilting and cation ordering are associated with a broad range of properties, including ferroelectric and photocatalytic behaviour in oxides and optoelectronic behaviour in halides~\cite{kojima2009,green2014,castelli2012,filip2018geometry}. This chemical flexibility makes perovskites attractive for materials design, but it also makes the search space large: a candidate must adopt the desired framework as well as exhibit the required electronic properties.

The accessible perovskite chemistry is not restricted to one cation at each crystallographic site. In mixed-B compositions, two or more cation species occupy the octahedral sublattice. In an ordered double perovskite, conventionally written as A$_2$BB$'$X$_6$, the two B-site species occupy distinct, often alternating, octahedral sublattices. Related ordered compositions can contain more complex site populations. Because the B--X octahedra form the connected framework, changing the cations assigned to these sites changes the local chemical environments that govern bonding and the electronic states near the band edges~\cite{wolf2021double}. A search restricted to a single-B prototype therefore samples only part of the structural and compositional diversity available to perovskites. Assessing this diversity requires site-resolved structural information; the total number of elements in a formula does not by itself identify which crystallographic sites they occupy.

A second difficulty is that composition does not determine topology. A nominal anion-to-cation ratio of 3:2 can also occur in structures that do not contain a three-dimensional network of corner-sharing BX$_6$ octahedra, including chain-like and layered arrangements documented across perovskite-derived chemistries~\cite{sopiha2022chalcogenide,goldschmidt1926,bartel2019tau}. Thus, a formula-based screen can admit structures outside the perovskite family. For a generative search, this distinction is especially important because the model proposes both atomic arrangements and compositions. Structural-family recovery must be assessed from the connectivity of the generated crystal, rather than inferred from stoichiometry or a tolerance factor alone.

The bandgap is a useful electronic descriptor for screening perovskites for optoelectronic applications. First-principles calculations can evaluate the gap for a specified crystal structure, but their computational cost limits exhaustive evaluation over a large set of generated candidates. In addition, semilocal density-functional approximations do not reproduce experimental bandgaps systematically; the Kohn--Sham gap from a generalized-gradient approximation is therefore not interchangeable with an optical or measured gap~\cite{perdew1983,sham1983,borlido2019}. In this work, 1.4 eV is the target supplied to MatterGen's $\texttt{dft\_band\_gap}$ checkpoint. It is a computational specification, not a claim about the optimum bandgap for a particular device. The downstream ALIGNN predictor is trained separately on the GGA/GGA+$U$ bandgap labels reported by the Materials Project (MP)~\cite{jain2013mp}; these MP labels are not the conditioning labels used to train MatterGen.

Generative models offer a route to exploring periodic crystals without prescribing a single prototype in advance. CDVAE and DiffCSP established approaches for generating or predicting periodic structures, while MatterGen jointly models atomic species, positions and lattice parameters and supports property-conditioned sampling through classifier-free guidance~\cite{xie2022cdvae,jiao2023diffcsp,zeni2025mattergen,ho2022cfg}. In parallel, crystal-graph models provide a means of estimating material properties from atomic structure~\cite{xie2018cgcnn,choudhary2021alignn}. Recent studies have compared generative crystal models with enumeration and substitution strategies~\cite{szymanski2025baselines}, demonstrated property-conditioned material generation~\cite{dobberstein2026materium}, and combined diffusion-based generation with first-principles assessment~\cite{mal2026diffcrysgen}. High-throughput calculations have also mapped the stability and electronic properties of perovskite compositions within enumerated chemical spaces~\cite{castelli2012,emery2017abo3}. These studies establish the potential of data-driven crystal exploration, but metrics such as general crystal validity or database novelty do not alone establish whether a generator recovers a particular structural family or explores its site-occupancy variants. Nor does a target-conditioned output guarantee that a downstream property model can rank the resulting structures reliably.

Here, we examine the use of the released MatterGen bandgap model to explore \ABX{} perovskites. The analysis is organized around three connected questions. First, to what extent do generated structures satisfy a three-dimensional corner-sharing perovskite definition, and what site-occupancy and anion chemistries do the accepted structures contain? Second, how are their predicted bandgaps distributed relative to the specified target, and how does the reliability of the ALIGNN predictor change between interpolation within known chemistry and a B-site-disjoint test? Third, do selected generated structures retain their perovskite connectivity after first-principles relaxation, and how do their calculated gaps compare with predictions made on the initial and relaxed geometries? A two-prototype substitution search is used as a reference for the structural space accessible to conventional enumeration. The comparison is not a matched search-space or compute-budget benchmark; its purpose is to identify which occupancy variants are excluded by the selected prototypes.

\section{Methods}

\begin{figure}[htbp]
  \centering
  \includegraphics[width=0.98\linewidth]{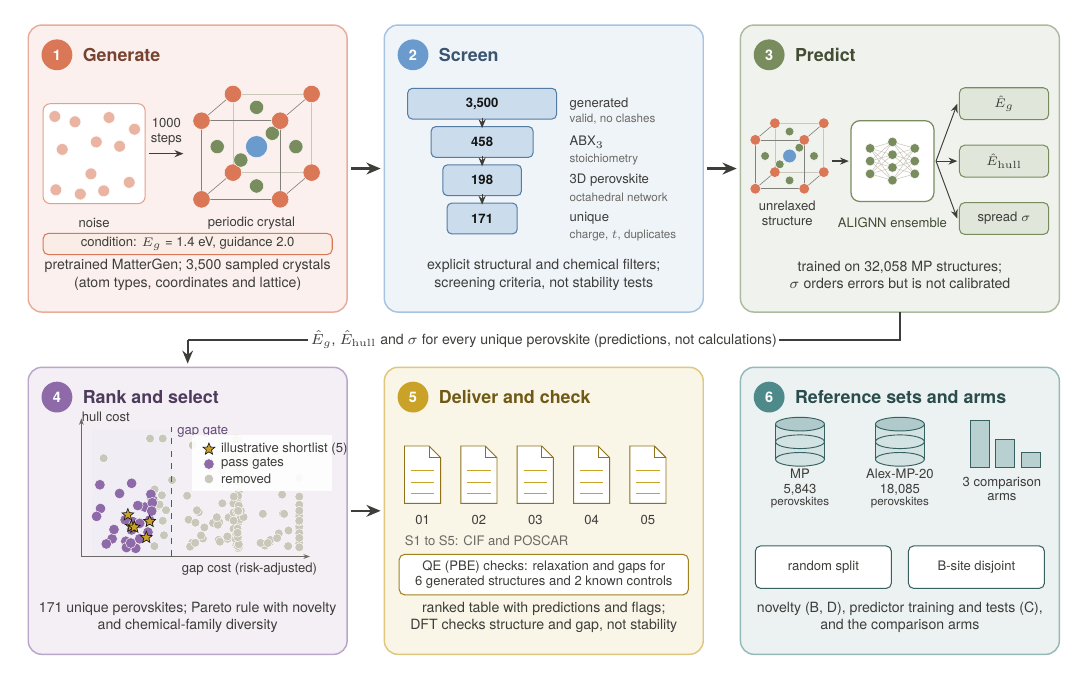}
  \caption{Overview of the structure-aware screening strategy. Periodic crystals are generated, classified using composition and octahedral connectivity, analysed for site occupancy and predicted bandgap, and assessed using selected first-principles calculations.}
  \label{fig:workflow}
\end{figure}

\subsection{Crystal generation and reference searches}\label{sec:gen}
Periodic crystals were generated with the released MatterGen model using its 1,000-step sampling procedure~\cite{zeni2025mattergen}. The released $\texttt{dft\_band\_gap}$ checkpoint was conditioned at 1.4 eV with classifier-free guidance strengths $\gamma=2$ and $\gamma=4$; the corresponding sets contained 3,500 and 800 structures. The conditioning property is associated with the Alex-MP-20 data release, whereas the downstream ALIGNN models were trained separately on curated MP GGA/GGA+$U$ labels. The two label sources were not merged. A further 1,200 structures were generated using the released unconditional checkpoint.

A prototype-substitution reference comprising 4,000 structures was constructed by inserting charge-balanced A--B--X combinations into the cubic $Pm\bar{3}m$ SrTiO$_3$ and orthorhombic Pnma CaTiO$_3$ prototypes. Compositions were filtered using a Goldschmidt tolerance-factor interval of 0.78--1.10 and cell volumes were scaled using Shannon ionic radii~\cite{shannon1976}. Because the reference uses only these two single-B prototypes, it cannot generate mixed-B or ordered double-perovskite structures. It is therefore used to compare accessible structural classes, not as a matched-budget measure of screening efficiency.

\subsection{Structural-family identification and database matching}\label{sec:topology}
A generated structure was considered compositionally compatible with the \ABX{} family when it was ordered, contained at least two cation species and had a 3:2 anion-to-cation ratio. This definition includes conventional \ABX{} cells and reduced compositions corresponding to ordered A$_2$BB$'$X$_6$ and related A$_3$B$_3$X$_9$-type structures. Compositionally compatible structures were then classified using their local coordination and connectivity. Cations with six anion neighbours within 1.25 times their shortest cation--anion distance were assigned as candidate B sites. Other cations were required to have at least eight anion neighbours within 1.3 times their shortest distance. Each anion had to bridge exactly two candidate B cations, and the resulting B--X network had to be periodic in all three dimensions, as determined from the rank of its periodic translations. Structures failing these conditions were not counted as three-dimensional corner-sharing perovskites. These thresholds define the operational classifier used here; they are not proposed as universal coordination boundaries for all distorted perovskites. B-site multiplicity was counted as the number of distinct elements assigned to the octahedral B sublattice by this criterion. It describes the explicit elemental arrangement in each generated structure and does not imply fractional occupancy or statistical disorder at a crystallographic site.

The accepted structures were additionally screened for charge balance using common oxidation states and for a tolerance factor between 0.78 and 1.10 where the required ionic radii and oxidation states were available. Structures for which a tolerance factor could not be evaluated were retained, rather than rejected on missing descriptor data. Duplicate structures were grouped using the pymatgen StructureMatcher, with primitive-cell reduction, volume scaling, $\texttt{ltol}=0.2$, $\texttt{stol}=0.3$ and an angular tolerance of 5$^\circ$~\cite{ong2013pymatgen}. Matching to MP and Alex-MP-20, the reference set associated with MatterGen training, was restricted to equal reduced formulas and applied to unrelaxed generated cells. The MP data were taken from the 2026.04.13 snapshot. The training pool was assembled from ordered ABC$_3$ entries and entries with up to 20 sites and an energy above the hull of at most 0.15 eV/atom; it was then restricted to ordered anion-containing structures with no more than 40 sites and both bandgap and hull labels. Duplicate removal left 32,058 structures. Database absence is therefore treated as an unmatched structure, not as proof of novelty.

\subsection{Bandgap prediction and validation}\label{sec:prediction}
Bandgaps and energies above the hull were estimated using ALIGNN ensembles~\cite{choudhary2021alignn}. The models were trained from scratch on 32,058 curated MP structures with both target labels. The ALIGNN architecture comprised four atomistic line-graph layers and four graph-convolution layers, with 256 features and an 8~\AA{} neighbour cutoff. Models were trained for 40 epochs with batch size 64 using AdamW~\cite{loshchilov2019adamw} and a one-cycle learning-rate schedule~\cite{smith2019superconvergence}; the final-epoch weights were retained. The principal property analysed here is the bandgap; predicted hull energies are model estimates and are not used as evidence of thermodynamic stability.

Predictor performance was evaluated using two formula-aware splits. In the random split, structures sharing a reduced formula were assigned to the same partition (25,807 training, 3,154 validation and 3,097 test structures), preventing formula-level leakage between partitions. In the B-site-disjoint split, structures containing any of 14 elements designated for the test (Be, Cd, Er, Ho, La, Os, Pa, Pu, Re, Rh, Tc, U, Yb and Zn) were excluded from training and validation wherever those elements occurred; the test set contained 290 perovskites whose B sites consisted only of these elements (24,153 training and 2,627 validation structures). The random-split ensemble contained three members and the B-site-disjoint ensemble two. Composition-based random-forest and $k$-nearest-neighbour models, as well as constant predictors, were used as baselines. Confidence intervals were estimated from 2,000 bootstrap resamples of test structures. Ensemble spread was evaluated as an error-ranking diagnostic, not interpreted as a calibrated predictive uncertainty.

\subsection{First-principles calculations}\label{sec:dft_methods}
Selected generated structures and two reference compounds were relaxed using Quantum ESPRESSO 7.3.1 with the PBE exchange--correlation functional~\cite{giannozzi2009qe,giannozzi2017qe}. Calculations used PSlibrary 1.0.0 pseudopotentials~\cite{dalcorso2014pslibrary}; an ultrasoft potential was used for Cs because the corresponding PAW file was not functional with the employed Quantum ESPRESSO version, and the Cd potential was taken from PSlibrary 0.3.1. The calculations were spin-unpolarized and scalar-relativistic. A Gaussian smearing width of 0.002 Ry, an SCF energy-convergence threshold of $10^{-9}$ Ry, wavefunction and charge-density cutoffs of 80 and 640 Ry, respectively, and variable-cell BFGS relaxation were used. The structures were standardized with spglib using a 0.05~\AA{} symmetry tolerance before relaxation~\cite{togo2024spglib}.

The DFT set comprised six generated structures and two controls (SrTiO$_3$ and cubic KNbO$_3$). The generated structures were selected from the available closed-shell examples and the candidate shortlist, including the largest generated cell that was computationally tractable; they do not constitute a random or statistically representative sample of the generated population. Open-shell and $4f$-element candidates were not included because the nonmagnetic calculations used here do not represent their magnetic ground states. Structural retention was assessed from the coordination criteria in \cref{sec:topology}: a structure was retained if each B cation remained six-coordinate, each X anion bridged two B cations, the B--X network remained three-dimensional and the A-site coordination remained at least eight. A structure failing the A-site or network criterion was classified as transformed. The relaxation protocol conserves the starting space group; it therefore does not test symmetry-lowering distortions or unstable phonon modes.

The bandgap was estimated from the lowest conduction-band and highest valence-band eigenvalues sampled on an even, $\Gamma$-centred mesh. The finite mesh provides a sampled estimate of the indirect gap and can miss band extrema between sampled $k$ points. Cutoffs and $k$-point spacing were checked for convergence on SrTiO$_3$, Cs$_2$NaRhBr$_6$ and Cs$_3$Mg$_2$CdBr$_9$. At 80 Ry, the total energy differed by less than 2.7 meV/atom from the 100 Ry result and the gap by less than 0.01 eV; production meshes changed gaps by up to 0.02 eV relative to denser meshes. No competing phases were calculated, so these calculations do not provide formation energies, hull distances or thermodynamic stability.

\section{Results and discussion}

\subsection{Recovery of the perovskite framework}\label{sec:res_yield}
The connectivity test substantially reduces the number of structures that qualify by composition alone. Among the 3,500 structures generated with the $\gamma=2$ conditioned model, 458 (13.1\%) satisfied the 3:2 anion-to-cation stoichiometric criterion, whereas 198 (5.7\% of all generated structures) also contained the required three-dimensional corner-sharing octahedral network. After chemical screening and duplicate removal, 171 unique perovskites remained. Thus, a composition-only filter would have admitted more than twice as many structures as the topology-based definition. This difference shows that nominal stoichiometry is insufficient for recovering the perovskite family from generated crystals.

The unconditional checkpoint produced 12 unique perovskites from 1,200 structures, compared with 171 from 3,500 structures in the $\gamma=2$ set. The observed unique-perovskite yields are 10.0 and 48.9 per 1,000 generated structures, respectively. This difference cannot be assigned to bandgap conditioning because the conditioned and unconditional samples come from different checkpoints, and the unconditional set contains only 12 accepted perovskites. The $\gamma=4$ set, generated from the same conditioned checkpoint as the $\gamma=2$ set, yielded 24 unique perovskites from 800 samples.

\begin{table}[htbp]
  \centering
  \caption{Structural and target-window yields for the three MatterGen sampling sets. The target-window fraction is calculated over unique accepted perovskites and uses predicted gaps within $1.4\pm0.3$ eV. Wilson 95\% confidence intervals are shown for the window fraction.}
  \label{tab:yield}
  \footnotesize
  \setlength{\tabcolsep}{3.0pt}
  \begin{tabular}{lrrrrr}
    \toprule
    Sampling set & $N_{\rm gen}$ & Unique & Yield (\%) & Multiple B (\%) & In window (\%)\\
    \midrule
    Unconditional & 1,200 & 12 & 1.0 & 75 & 16.7 (4.7--44.8)\\
    Conditioned, $\gamma=2$ & 3,500 & 171 & 4.9 & 69 & 8.2 (4.9--13.3)\\
    Conditioned, $\gamma=4$ & 800 & 24 & 3.0 & 71 & 25.0 (12.0--44.9)\\
    \bottomrule
  \end{tabular}
\end{table}

\subsection{B-site occupancy and chemical diversity}\label{sec:res_chem}
The clearest difference between crystal generation and the two-prototype substitution reference is the range of B-site arrangements. Of the 171 accepted generated perovskites, 118 (69\%) contain more than one B-site element. Approximately two thirds of the generated halide perovskites are ordered A$_2$BB$'$X$_6$ structures, including compositions such as Cs$_2$AgSbCl$_6$ and Rb$_2$CuBiBr$_6$. The substitution reference, by construction, contains only one B-site species and cannot represent these configurations. This comparison therefore demonstrates a difference in accessible structural chemistry, not a higher efficiency of the generator: the two approaches do not search the same compositional or structural space, and their computational budgets are not matched.

The generated set also differs from the MP perovskite reference set in its anion distribution. Chloride-, bromide- and iodide-containing compositions account for 45\% of generated perovskites, compared with 22\% in MP; oxides account for 26\% compared with 62\%, while fluorides account for 26\% compared with 15\%. The A site is dominated by Cs, Rb and K. All 49 elements assigned to generated B sites are already represented in the MP perovskite reference set. The result is therefore recombination of known elements into different compositions and site arrangements, rather than access to previously unrepresented B-site elements. The observed distributions describe the generated output; they cannot be attributed specifically to bandgap conditioning because the conditioned and unconditional arms use different checkpoints.

\begin{figure}[htbp]
  \centering
  \includegraphics[width=0.98\linewidth]{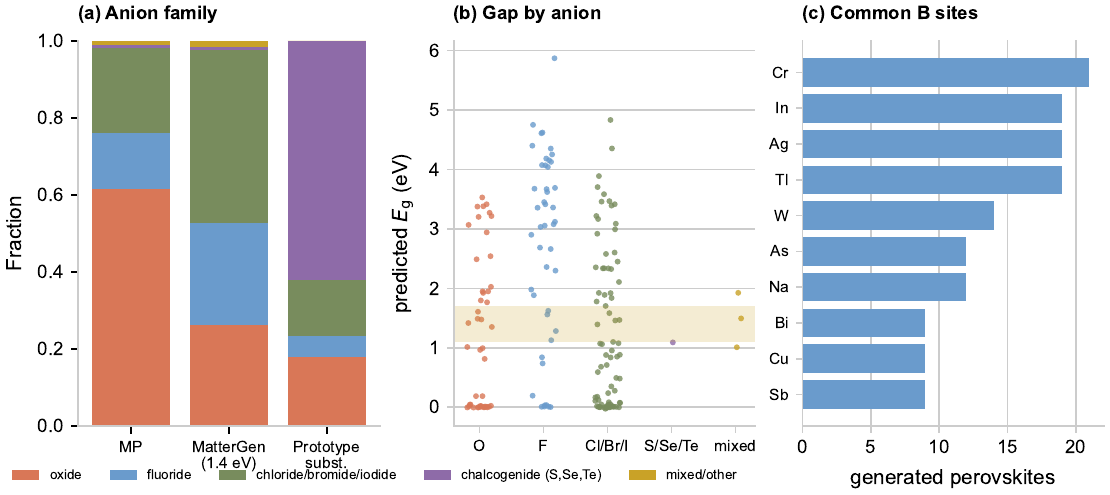}
  \caption{Chemical features of the accepted generated perovskites compared with the MP reference set. (a) Fractions containing chlorides, bromides or iodides, oxides and fluorides. (b) Fraction with more than one B-site element in the generated set and in the two-prototype substitution reference. These are observed distributions, not a controlled estimate of the effect of bandgap conditioning.}
  \label{fig:chemistry}
\end{figure}

The distinction between site occupancy and total compositional complexity is important. Although 69\% of generated structures have multiple B-site elements, 64\% contain four or more elements overall, close to the 66\% value in MP. Total element count does not identify crystallographic site assignment: structures with the same number of elements can differ in whether chemical disorder or ordering occurs on the A, B or X sublattice. The high mixed-B fraction is therefore a site-resolved feature of the generated structures and should not be inferred from formula complexity alone. \Cref{fig:chemistry} summarizes the differences in anion chemistry and B-site multiplicity.

A comparison with MP and Alex-MP-20 further distinguishes chemical recombination from structural novelty. Of the 171 generated perovskites, 101 match an MP structure and 125 match a structure in Alex-MP-20; 134 match at least one of the two sets, while 37 match neither. The unmatched fraction is not a novelty yield. Matching was performed on unrelaxed generated cells and can miss structures that become equivalent after relaxation or standardization, and the search covers only these two databases. None of the nine structures meeting the study's predicted-gap, predicted-hull and ensemble-spread criteria is unmatched from both reference sets. The present data therefore support exploration of varied site arrangements, but do not establish the discovery of a new stable perovskite.

A more restrictive screen required all three predicted criteria to be met: $|\hat E_{\mathrm{g}}-1.4|\leq0.3$ eV, predicted $\hat E_{\mathrm{hull}}\leq0.10$ eV/atom and gap-ensemble spread $\sigma_{\mathrm{ens}}\leq0.30$ eV. The resulting pass rates are listed in \cref{tab:combined_screen}. The prototype-substitution reference gives the highest rate, 22.8 screen-qualified structures per 1,000 constructed structures, compared with 2.6 for $\gamma=2$, 7.5 for $\gamma=4$ and 0.8 for unconditional generation. This is a model-filter yield, not a stable-material yield: the hull values are ALIGNN predictions, and the ensemble spread is not calibrated. It is also not a compute-normalized comparison. The substitution pool was filtered by charge balance and tolerance factor before the 4,000 structures were built, and the two approaches explore different structural spaces. Its chemical distribution also differs substantially: 62\% of the substitution structures are chalcogenides, compared with 0.6\% of the generated perovskites; chalcogenides are sparsely represented in the training data and their predictions are less certain. Thus, the prototype reference is more productive under this particular screen within its restricted search space, whereas crystal generation accesses mixed-B and ordered configurations that the reference cannot represent.

\begin{table}[htbp]
  \centering
  \caption{Yield of structures passing the combined predicted-property screen: $|\hat E_{\mathrm{g}}-1.4|\leq0.3$ eV, $\hat E_{\mathrm{hull}}\leq0.10$ eV/atom and $\sigma_{\mathrm{ens}}\leq0.30$ eV. Rates are normalized by 1,000 generated or constructed structures. The hull and spread values are model estimates; these rates are not stable-material yields.}
  \label{tab:combined_screen}
  \small
  \begin{tabular}{lrr}
    \toprule
    Search set & Structures & Screen-qualified per 1,000 \\
    \midrule
    Unconditional generation & 1,200 & 0.8 \\
    Conditioned generation ($\gamma=2$) & 3,500 & 2.6 \\
    Conditioned generation ($\gamma=4$) & 800 & 7.5 \\
    Prototype substitution & 4,000 & 22.8 \\
    \bottomrule
  \end{tabular}
\end{table}

\subsection{Bandgap distribution and target-window screening}\label{sec:res_window}
The median ALIGNN-predicted bandgap of the $\gamma=2$ set is 1.48 eV, closer to the 1.4 eV specification than the 0.78 eV median of the unconditional set. The corresponding MP reference median is 0.69 eV. Fourteen of the 171 $\gamma=2$ perovskites (8.2\%) lie within 1.1--1.7 eV, compared with 7.3\% of the MP reference, 2 of 12 (16.7\%) in the unconditional set and 6 of 24 (25\%) in the $\gamma=4$ set. The comparison between the unconditional and $\gamma=2$ outputs is not statistically conclusive (Fisher's exact test, $p=0.28$), and the unconditional estimate is based on only 12 unique structures.

These data do not show that the conditioned checkpoint improves target-window success. They instead show that the $\gamma=2$ output has a median closer to the specified value while retaining a target-window fraction close to the MP reference. The difference between a distributional shift and target-window enrichment is important for inverse design: moving the centre of a predicted-property distribution towards a target does not necessarily increase the number of candidates that satisfy a finite acceptance interval. The interpretation is further limited by the chemical shift towards heavier halides and by the observed tendency of ALIGNN to predict higher gaps than PBE for the small subset of generated halides assessed below. The present comparisons do not separate generator effects from predictor bias.

\subsection{Predictor accuracy and chemical transferability}\label{sec:res_pred}
On the formula-grouped random test split, the ALIGNN ensemble predicts bandgaps with a mean absolute error (MAE) of 0.25 eV ($n=576$), compared with 0.58 eV for the composition-based random forest. The corresponding $R^2$ of ALIGNN is 0.85. This result indicates that structural information improves interpolation within the chemical distribution represented in the training data. It does not, by itself, demonstrate reliable prediction for compositions or local environments that are poorly represented in that data.

The B-site-disjoint test exposes this limitation. On this test, the ALIGNN MAE is 0.55 eV and $R^2=0.05$, while a constant-zero predictor gives an MAE of 0.53 eV. ALIGNN therefore does not improve on the constant baseline for this particular chemical-shift test. The test is deliberately severe: 75\% of its labels are zero and 53\% of its structures contain lanthanide or actinide elements. Only 13 of the 171 generated perovskites contain any of the 14 held-out elements at the B site. This split is consequently a stress test of transfer to unfamiliar B-site chemistry, not an estimate of the expected error across the generated population. It nevertheless shows that good random-split accuracy should not be taken as evidence of broad chemical transferability. Recent analysis of machine-learned bandgaps has likewise shown that zero-gap-rich datasets can obscure regression errors and that chemistry-grouped splits distinguish interpolation from transfer to held-out chemical families~\cite{solecki2026bandgaps}. The restricted MAE values reported in \cref{tab:baselines} provide an additional view over the 0.5--3 eV interval, although this interval is not identical to a full non-zero-gap evaluation.

The ensemble spread has a positive rank correlation with absolute error on the random split (Spearman $\rho=0.67$), but the corresponding prediction intervals are not calibrated. The ensemble contains only three members for the random split and two for the B-site-disjoint split, so the spread should be treated as a qualitative error-ranking signal rather than a probabilistic uncertainty. These limitations matter for bandgap-targeted screening: the predicted values can help organize candidates within represented chemistry, but the spread does not supply a calibrated confidence interval, and candidates in unfamiliar chemical environments require independent calculations.

\begin{table}[htbp]
  \centering
  \caption{Bandgap prediction on held-out MP perovskites. MAE values are in eV; intervals are 95\% bootstrap intervals. The restricted MAE is calculated for reference labels between 0.5 and 3 eV.}
  \label{tab:baselines}
  \footnotesize
  \setlength{\tabcolsep}{4.1pt}
  \begin{tabular}{lcccccc}
    \toprule
    & \multicolumn{3}{c}{Random split ($n=576$)} & \multicolumn{3}{c}{B-site-disjoint ($n=290$)}\\
    \cmidrule(lr){2-4}\cmidrule(lr){5-7}
    Predictor & MAE [95\% CI] & $R^2$ & MAE, 0.5--3 eV & MAE [95\% CI] & $R^2$ & MAE, 0.5--3 eV\\
    \midrule
    ALIGNN ensemble & 0.25 [0.20, 0.30] & 0.85 & 0.28 & 0.55 [0.43, 0.67] & 0.05 & 1.33\\
    Composition RF & 0.58 [0.53, 0.63] & 0.76 & 0.46 & 1.08 [0.98, 1.18] & -0.37 & 0.75\\
    Composition $k$NN & 1.12 [1.04, 1.20] & 0.29 & 0.81 & 1.41 [1.27, 1.57] & -1.48 & 0.87\\
    Training median & 1.43 [1.33, 1.54] & -0.23 & 1.07 & 0.91 [0.83, 1.01] & -0.02 & 0.66\\
    Constant zero & 1.52 [1.37, 1.67] & -0.74 & 1.73 & 0.53 [0.40, 0.68] & -0.19 & 1.35\\
    \bottomrule
  \end{tabular}
\end{table}

\begin{figure}[htbp]
  \centering
  \includegraphics[width=\linewidth]{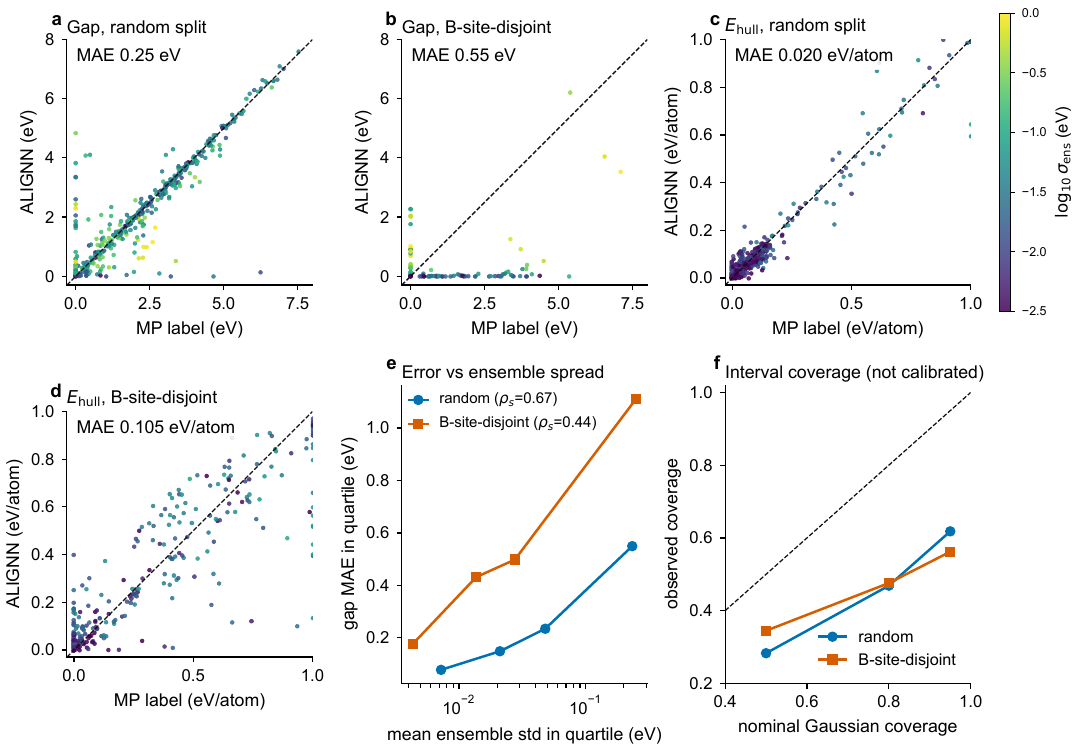}
  \caption{ALIGNN predictions for held-out MP perovskites. Panels (a)--(d) compare ensemble predictions with reference labels for bandgap and energy above the hull under random and B-site-disjoint splits. Panels (e) and (f) show the relationship between ensemble spread, absolute error and nominal interval coverage. The B-site-disjoint bandgap model does not outperform a constant-zero predictor, and intervals constructed from ensemble spread are not calibrated.}
  \label{fig:predictors}
\end{figure}

The ensemble spread orders errors (Spearman $\rho=0.67$ for the random split and $0.44$ for the B-site-disjoint split), but the correlations are not directly comparable because the ensembles differ in size. For the 101 generated structures matching an MP entry, the prediction on the generated geometry differs from the MP label by an MAE of 0.16 eV. Because these compounds are represented in the training data and the MP labels refer to relaxed structures, this comparison measures sensitivity to geometry rather than out-of-sample accuracy. The largest deviations include open-shell compounds for which the MP labels depend on magnetic order and $+U$ settings.

\subsection{First-principles response of generated structures}\label{sec:dft}
Eight structures were relaxed using variable-cell PBE calculations: six generated structures and the SrTiO$_3$ and KNbO$_3$ controls. Seven retained the defined corner-sharing framework after relaxation. This retention should be interpreted in the context of the imposed symmetry: the calculations conserve the starting space group and therefore do not test symmetry-lowering octahedral tilts or other distortions that require a larger structural search. Retention under this protocol is not evidence of thermodynamic or dynamical stability.

TlPbCl$_3$ was the exception under the adopted structural classification. During relaxation its cell volume increased by 3.29\%, its mean B--X distance increased from 2.84 to 2.89~\AA, and the smallest B--X--B angle decreased to 166$^\circ$. The B coordination and bridging remained satisfied, but the A-site coordination decreased from 12 to 5, below the minimum used in the perovskite-family definition. The structure was therefore classified as transformed. This result shows that checking the octahedral network alone is not sufficient: the coordination of the nominal A site can also determine whether a relaxed structure remains within the adopted perovskite definition.

\begin{table}[htbp]
  \centering
  \caption{Structural metrics before and after variable-cell PBE relaxation. The retention class follows the coordination and connectivity criteria in \cref{sec:dft_methods}. The relaxation conserves the starting space group; this table does not assess thermodynamic or dynamical stability.}
  \label{tab:retention}
  \scriptsize
  \setlength{\tabcolsep}{2.5pt}
  \begin{tabular}{llcccccccll}
    \toprule
    ID & Formula & Atoms & SG (start) & SG (relaxed) & $\Delta V$ (\%) & $\langle d_{BX}\rangle$ (\AA) & min. B--X--B ($^\circ$) & A CN & Class & Database match\\
    \midrule
    D1 & SrTiO$_3$ & 5 & $Pm\bar{3}m$ & $Pm\bar{3}m$ & -0.36 & 1.971 & 180 & 12$\to$12 & Retained & MP\\
    D2 & KNbO$_3$ & 5 & $Pm\bar{3}m$ & $Pm\bar{3}m$ & -2.26 & 2.015 & 180 & 12$\to$12 & Retained & MP\\
    D3 & Cs$_2$NaRhBr$_6$ & 10 & $Fm\bar{3}m$ & $Fm\bar{3}m$ & -0.81 & 2.759 & 180 & 12$\to$12 & Retained & Alex-MP-20\\
    D4 & Tl$_3$AsCl$_6$ & 10 & $Fm\bar{3}m$ & $Fm\bar{3}m$ & -0.55 & 2.750 & 180 & 12$\to$12 & Retained & Neither\\
    D5 & TlPbCl$_3$ & 5 & Amm2 & Amm2 & +3.29 & 2.890 & 166 & 12$\to$5 & Transformed & Alex-MP-20\\
    D6 & Cs$_3$Mg$_2$CdBr$_9$ & 15 & $P\bar{3}m1$ & $P\bar{3}m1$ & +0.39 & 2.751 & 180 & 12$\to$12 & Retained & Alex-MP-20\\
    D7 & Tl$_2$CuBiCl$_6$ & 10 & $Fm\bar{3}m$ & $Fm\bar{3}m$ & +0.34 & 2.617 & 180 & 12$\to$12 & Retained & Neither\\
    D8 & Cs$_2$LiAsI$_6$ & 10 & $Fm\bar{3}m$ & $Fm\bar{3}m$ & -0.18 & 2.962 & 180 & 12$\to$12 & Retained & Alex-MP-20\\
    \bottomrule
  \end{tabular}
\end{table}

The comparison between ALIGNN and PBE depends on whether the model is evaluated on the initial or relaxed geometry. On the initial structures, the absolute differences range from 0.13 to 0.60 eV, with a mean of 0.31 eV; four of the eight structures are within the random-split MAE of 0.25 eV. When ALIGNN is evaluated on the relaxed geometries, the differences range from 0.04 to 0.43 eV. For the five generated halides whose frameworks were retained, ALIGNN predicts a higher gap than PBE in every case, with a mean difference of 0.30 eV. This consistent sign is evidence of a possible bias for this small evaluated subset, but the selected structures are too few and too chemically restricted to establish a general bias for halide perovskites.

TlPbCl$_3$ illustrates the effect of geometry on the comparison. ALIGNN predicts 1.92 eV on the generated starting structure, whereas PBE gives 2.53 eV after relaxation. On the relaxed geometry, ALIGNN predicts 2.67 eV, reducing the difference from 0.60 to 0.14 eV in magnitude and reversing its sign. Thus, a substantial part of the initial discrepancy is associated with the structural change during relaxation, rather than with the predictor alone.

Three of the six generated structures have PBE gaps between 1.1 and 1.7 eV: Tl$_3$AsCl$_6$ (1.46 eV), Cs$_3$Mg$_2$CdBr$_9$ (1.69 eV) and Cs$_2$LiAsI$_6$ (1.34 eV). These structures were selected for computational feasibility or shortlist membership, not because their predicted gaps fell within the target window. In fact, ALIGNN placed all three outside the interval on their starting geometries. None of the generated structures in this DFT subset was simultaneously inside the target window by both ALIGNN and PBE. The calculations therefore do not validate successful target-window selection; instead, they show that geometry relaxation and predictor bias can change which generated structures appear promising. The eight calculations are a selected case study and do not provide a statistical estimate of predictor error over the generated population.

The electronic-structure comparison has further limitations. The generated DFT set excludes open-shell candidates, and the spin-unpolarized calculations do not address magnetic ordering or magnetic contributions to the bandgap. Spin--orbit coupling was not included, although it can be important for compounds containing heavy elements such as Tl, Bi and I. The calculated PBE gaps should therefore be read as a consistent first-principles assessment within the present approximation, not as predictions of optical gaps or experimental performance.

\begin{table}[htbp]
  \centering
  \caption{ALIGNN bandgap predictions and PBE bandgaps. Starting- and relaxed-geometry predictions are reported separately. The PBE gap is the sampled indirect gap; the minimum direct gap on the same mesh is also shown. $\Delta$ is ALIGNN minus PBE.}
  \label{tab:dftgap}
  \scriptsize
  \setlength{\tabcolsep}{3.6pt}
  \begin{tabular}{llcccccccc}
    \toprule
    ID & Formula & MP label & ALIGNN start & ALIGNN relaxed & PBE gap & PBE min. direct & $\Delta_{\rm start}$ & $\Delta_{\rm relaxed}$\\
    \midrule
    D1 & SrTiO$_3$ & 1.77 & $1.61\pm0.13$ & 1.62 & 1.83 & 2.18 & -0.22 & -0.20\\
    D2 & KNbO$_3$ & 1.42 & $1.35\pm0.09$ & 1.48 & 1.52 & 2.16 & -0.17 & -0.04\\
    D3 & Cs$_2$NaRhBr$_6$ & -- & $1.08\pm0.09$ & 1.06 & 0.94 & 0.94 & +0.13 & +0.12\\
    D4 & Tl$_3$AsCl$_6$ & -- & $1.84\pm0.02$ & 1.84 & 1.46 & 1.46 & +0.38 & +0.39\\
    D5 & TlPbCl$_3$ & -- & $1.92\pm0.08$ & 2.67 & 2.53 & 2.53 & -0.60 & +0.14\\
    D6 & Cs$_3$Mg$_2$CdBr$_9$ & -- & $2.05\pm0.18$ & 2.06 & 1.69 & 1.85 & +0.37 & +0.37\\
    D7 & Tl$_2$CuBiCl$_6$ & -- & $1.10\pm0.08$ & 1.12 & 0.90 & 1.70 & +0.20 & +0.22\\
    D8 & Cs$_2$LiAsI$_6$ & -- & $1.78\pm0.03$ & 1.77 & 1.34 & 1.68 & +0.43 & +0.43\\
    \bottomrule
  \end{tabular}
\end{table}

\begin{figure}[htbp]
  \centering
  \includegraphics[width=0.98\linewidth]{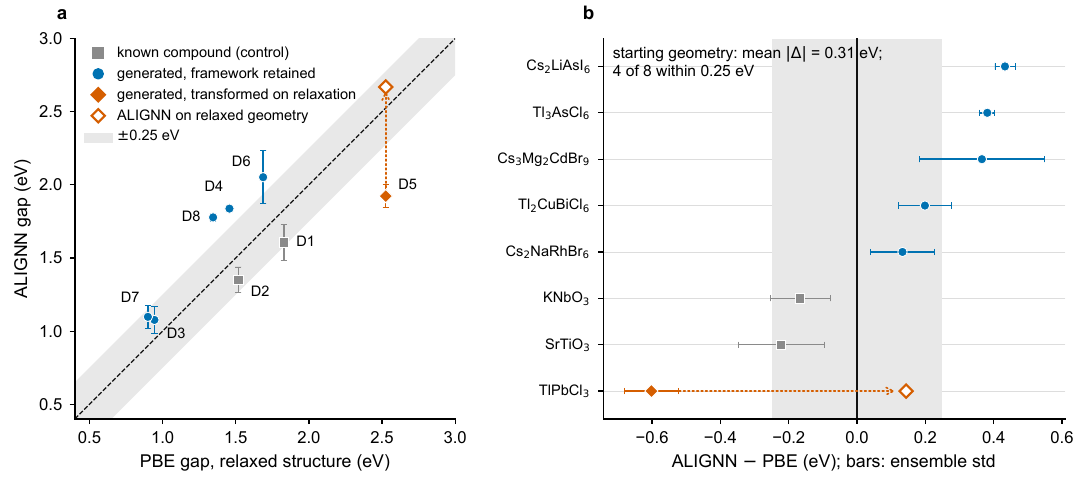}
  \caption{ALIGNN predictions compared with PBE bandgaps for the two controls and six generated structures. (a) Predictions on the starting geometries versus PBE gaps of the relaxed structures; the open marker for TlPbCl$_3$ shows the ALIGNN prediction on its relaxed geometry. (b) Difference between ALIGNN and PBE. The structures were selected for computational feasibility or shortlist membership and are not a random sample of generated perovskites.}
  \label{fig:dftgaps}
\end{figure}

\begin{figure}[htbp]
  \centering
  \includegraphics[width=0.98\linewidth]{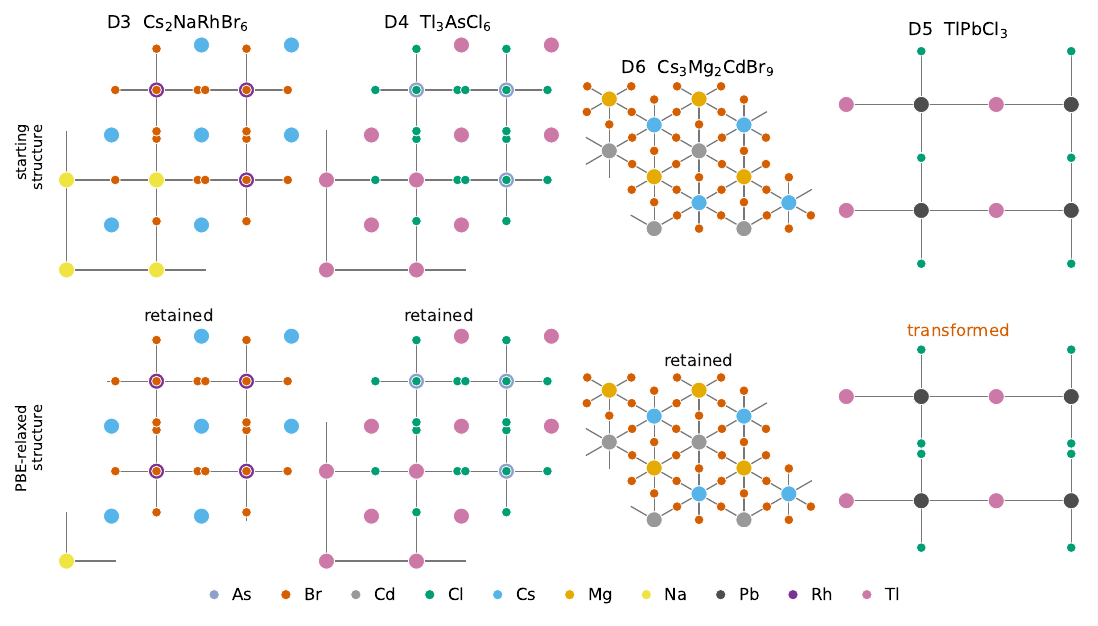}
  \caption{Starting and PBE-relaxed structures for selected generated compounds. The projections show the standardized starting structures and the corresponding relaxed cells for Cs$_2$NaRhBr$_6$, Tl$_3$AsCl$_6$, Cs$_3$Mg$_2$CdBr$_9$ and TlPbCl$_3$. The first three retain the defined corner-sharing framework, whereas TlPbCl$_3$ fails the A-site coordination criterion after relaxation.}
  \label{fig:dftstructures}
\end{figure}

\subsection{Relation to published crystal-generation benchmarks}\label{sec:benchmarks}
Published crystal-generation studies evaluate different stages of the discovery process. MatterGen reports structural uniqueness and novelty relative to a large reference set, while DiffCrysGen additionally assesses energy-based stability, relaxation fidelity and dynamical stability~\cite{zeni2025mattergen,mal2026diffcrysgen}. Materium evaluates three-dimensional structure generation and property controllability~\cite{dobberstein2026materium}. The present study instead focuses on recovery of a specific perovskite topology, site-resolved B-site occupancy and the reliability of bandgap screening. These evaluations are complementary; their metrics should not be ranked unless their definitions, reference databases and evaluation budgets are aligned.

A closer comparison is possible for target-gap screening. For the 3 eV task studied by Szymanski and Bartel, the fraction of novel candidates within $\pm0.5$ eV of the target was 11.2\% for random enumeration, 21.4\% after CGCNN filtering, 37.2\% for ion exchange and 61.4\% for FTCP~\cite{szymanski2025baselines}. The present fraction is 8.2\% within $1.4\pm0.3$ eV among 171 topology-filtered perovskites, using ALIGNN-predicted gaps. The target, interval, candidate family, novelty criterion and property-evaluation method differ, so these values do not support a numerical ranking. They do show why target-window yield should be reported alongside structural recovery and why conditioned generation should be compared with property-screening baselines as well as unconditional sampling.

\begin{table}[!htbp]
  \centering
  \caption{Context for the present perovskite-specific evaluation relative to recent crystal-generation studies. Metrics are reported as defined in each source and are not directly comparable unless the evaluation protocols are aligned.}
  \label{tab:literature}
  \footnotesize
  \setlength{\tabcolsep}{3.2pt}
  \renewcommand{\arraystretch}{1.0}
  \begin{tabularx}{\linewidth}{>{\RaggedRight\arraybackslash}p{0.16\linewidth}>{\RaggedRight\arraybackslash}p{0.43\linewidth}>{\RaggedRight\arraybackslash}X}
    \toprule
    Study & Reported evaluation & Relation to the present study\\
    \midrule
    MatterGen~\cite{zeni2025mattergen} & Reports 100\% unique structures among 1,000 samples and 61\% novel structures relative to Alex-MP-ICSD, alongside stability and relaxation metrics. & The present 22\% database-unmatched fraction uses MP and Alex-MP-20, equal-formula matching and unrelaxed cells; it is not equivalent to the published novelty metric.\\
    Szymanski and Bartel~\cite{szymanski2025baselines} & Compares enumeration, ML filtering, ion exchange and generative models. For 3 eV targeting, the reported $\pm0.5$ eV hit fractions are 11.2\%, 21.4\%, 37.2\% and 61.4\% for random enumeration, CGCNN filtering, ion exchange and FTCP, respectively. & The present 8.2\% is for a 1.4$\pm$0.3 eV interval, topology-filtered perovskites and ALIGNN-predicted gaps; the protocols do not support direct ranking.\\
    Materium~\cite{dobberstein2026materium} & Evaluates autoregressive three-dimensional crystal generation and property controllability; it reports an 85.4\% structure-match rate. & Structure matching is not equivalent to the present perovskite-topology recovery or target-window fraction.\\
    DiffCrysGen~\cite{mal2026diffcrysgen} & Reports validity, formation-energy and stability metrics, relaxation, phonon assessment and generation throughput (100,000 structures in 5.41 min on one H100 GPU). & The present study has no matched throughput benchmark, competing-phase stability or phonon calculations; its six generated DFT cases cannot support those comparisons.\\
    Present study & Measures topology-positive perovskite recovery, site-resolved B-site multiplicity, target-window fraction and ALIGNN--PBE agreement for selected relaxed structures. & Provides family-specific evidence on structural and chemical coverage and downstream screening limits, not superiority in general validity, stability, novelty or throughput.\\
    \bottomrule
  \end{tabularx}
\end{table}

\FloatBarrier
\section{Conclusions}
This study examines how a pretrained crystal-diffusion model explores the structural and chemical space of \ABX{} perovskites and how reliably the resulting structures can be screened for a specified bandgap. A connectivity-based definition is essential: of the 458 generated structures satisfying the nominal 3:2 anion-to-cation stoichiometric criterion, 198 contain the required three-dimensional corner-sharing network, leaving 171 unique perovskites after chemical screening and duplicate removal. The accepted set includes substantial B-site multiplicity, with 69\% containing more than one B-site element. This is the principal difference from the two-prototype substitution reference, which is restricted to single-B structures. The generated structures recombine elements already represented in MP rather than introducing previously absent B-site chemistry.

The $\gamma=2$ set has a median predicted gap closer to the 1.4 eV specification than the unconditional set, while its target-window fraction (8.2\%) is close to the MP reference fraction (7.3\%). The ALIGNN model performs well on a formula-grouped random split but does not outperform a constant predictor on the B-site-disjoint test, demonstrating limited transferability to unfamiliar B-site chemistry. The selected PBE calculations further show that relaxation can change the structural classification and that starting-geometry bandgap predictions can differ from PBE values by several tenths of an electronvolt. In this subset, no generated structure is simultaneously within the target interval by both ALIGNN and PBE.

The results support crystal diffusion as a way to explore mixed-site and ordered perovskite configurations beyond a narrow prototype set, but they do not establish new stable compounds or successful inverse design at the specified gap. The released generator also limits the size of the sampled unit cells, restricting access to larger-period ordering patterns. Stronger claims would require replicated guidance-strength comparisons, including a $\gamma=0$ sample from the conditioned checkpoint, a chemically and computationally matched baseline, validation and sensitivity analysis of the topology classifier, broader first-principles assessment including magnetic and spin--orbit effects, and stability calculations against competing phases and phonons. Independent assessment also requires release of the code, generated structures, model provenance, processed data and calculation inputs and outputs in a reproducible form.

\section*{Declaration of generative AI use}
ChatGPT and Claude were used to assist with code audit, and revising the language of the manuscript. The authors reviewed the generated material and take responsibility for the final content.

\section*{Code and data availability}
Code and supporting data are intended to be available at \url{https://github.com/sunraj-patel/Diffusion-Generative-Screening}.

\bibliographystyle{unsrtnat}
\bibliography{refs}

\begin{thebibliography}{31}
\providecommand{\natexlab}[1]{#1}
\providecommand{\url}[1]{\texttt{#1}}
\expandafter\ifx\csname urlstyle\endcsname\relax
  \providecommand{\doi}[1]{doi: #1}\else
  \providecommand{\doi}{doi: \begingroup \urlstyle{rm}\Url}\fi

\bibitem[Kojima et~al.(2009)Kojima, Teshima, Shirai, and Miyasaka]{kojima2009}
Akihiro Kojima, Kenjiro Teshima, Yasuo Shirai, and Tsutomu Miyasaka.
\newblock Organometal halide perovskites as visible-light sensitizers for
  photovoltaic cells.
\newblock \emph{Journal of the American Chemical Society}, 131:\penalty0
  6050--6051, 2009.
\newblock \doi{10.1021/ja809598r}.

\bibitem[Green et~al.(2014)Green, Ho-Baillie, and Snaith]{green2014}
Martin~A. Green, Anita Ho-Baillie, and Henry~J. Snaith.
\newblock The emergence of perovskite solar cells.
\newblock \emph{Nature Photonics}, 8:\penalty0 506--514, 2014.
\newblock \doi{10.1038/nphoton.2014.134}.

\bibitem[Castelli et~al.(2012)Castelli, Landis, Thygesen, Dahl, Chorkendorff,
  Jaramillo, and Jacobsen]{castelli2012}
Ivano~E. Castelli, David~D. Landis, Kristian~S. Thygesen, Søren Dahl,
  Ib~Chorkendorff, Thomas~F. Jaramillo, and Karsten~W. Jacobsen.
\newblock New cubic perovskites for one- and two-photon water splitting using
  the computational materials repository.
\newblock \emph{Energy and Environmental Science}, 5:\penalty0 9034, 2012.
\newblock \doi{10.1039/C2EE22341D}.

\bibitem[Filip and Giustino(2018)]{filip2018geometry}
Marina~R. Filip and Feliciano Giustino.
\newblock The geometric blueprint of perovskites.
\newblock \emph{Proceedings of the National Academy of Sciences}, 115:\penalty0
  5397--5402, 2018.
\newblock \doi{10.1073/pnas.1719179115}.

\bibitem[Wolf et~al.(2021)Wolf, Connor, Slavney, and
  Karunadasa]{wolf2021double}
Nicholas~R. Wolf, Benjamin~A. Connor, Adam~H. Slavney, and Hemamala~I.
  Karunadasa.
\newblock Doubling the stakes: The promise of halide double perovskites.
\newblock \emph{Angewandte Chemie International Edition}, 60:\penalty0
  16264--16278, 2021.
\newblock \doi{10.1002/anie.202016185}.

\bibitem[Sopiha et~al.(2022)Sopiha, Comparotto, Márquez, and
  Scragg]{sopiha2022chalcogenide}
K.~V. Sopiha, C.~Comparotto, J.~A. Márquez, and J.~J.~S. Scragg.
\newblock Chalcogenide perovskites: Tantalizing prospects, challenging
  materials.
\newblock \emph{Advanced Optical Materials}, 10:\penalty0 2101704, 2022.
\newblock \doi{10.1002/adom.202101704}.

\bibitem[Goldschmidt(1926)]{goldschmidt1926}
V.~M. Goldschmidt.
\newblock Die gesetze der krystallochemie.
\newblock \emph{Die Naturwissenschaften}, 14:\penalty0 477--485, 1926.
\newblock \doi{10.1007/BF01507527}.

\bibitem[Bartel et~al.(2019)Bartel, Sutton, Goldsmith, Ouyang, Musgrave,
  Ghiringhelli, and Scheffler]{bartel2019tau}
Christopher~J. Bartel, Christopher Sutton, Bryan~R. Goldsmith, Runhai Ouyang,
  Charles~B. Musgrave, Luca~M. Ghiringhelli, and Matthias Scheffler.
\newblock New tolerance factor to predict the stability of perovskite oxides
  and halides.
\newblock \emph{Science Advances}, 5:\penalty0 eaav0693, 2019.
\newblock \doi{10.1126/sciadv.aav0693}.

\bibitem[Perdew and Levy(1983)]{perdew1983}
John~P. Perdew and Mel Levy.
\newblock Physical content of the exact kohn-sham orbital energies: Band gaps
  and derivative discontinuities.
\newblock \emph{Physical Review Letters}, 51:\penalty0 1884--1887, 1983.
\newblock \doi{10.1103/PhysRevLett.51.1884}.

\bibitem[Sham and Schlüter(1983)]{sham1983}
L.~J. Sham and M.~Schlüter.
\newblock Density-functional theory of the energy gap.
\newblock \emph{Physical Review Letters}, 51:\penalty0 1888--1891, 1983.
\newblock \doi{10.1103/PhysRevLett.51.1888}.

\bibitem[Borlido et~al.(2019)Borlido, Aull, Huran, Tran, Marques, and
  Botti]{borlido2019}
Pedro Borlido, Thorsten Aull, Ahmad~W. Huran, Fabien Tran, Miguel A.~L.
  Marques, and Silvana Botti.
\newblock Large-scale benchmark of exchange-correlation functionals for the
  determination of electronic band gaps of solids.
\newblock \emph{Journal of Chemical Theory and Computation}, 15:\penalty0
  5069--5079, 2019.
\newblock \doi{10.1021/acs.jctc.9b00322}.

\bibitem[Jain et~al.(2013)Jain, Ong, Hautier, Chen, Richards, Dacek, Cholia,
  Gunter, Skinner, Ceder, and Persson]{jain2013mp}
Anubhav Jain, Shyue~Ping Ong, Geoffroy Hautier, Wei Chen, William~D. Richards,
  Stephen Dacek, Shreyas Cholia, Dan Gunter, David Skinner, Gerbrand Ceder, and
  Kristin~A. Persson.
\newblock Commentary: The materials project: A materials genome approach to
  accelerating materials innovation.
\newblock \emph{APL Materials}, 1:\penalty0 011002, 2013.
\newblock \doi{10.1063/1.4812323}.

\bibitem[Xie et~al.(2022)Xie, Fu, Ganea, Barzilay, and Jaakkola]{xie2022cdvae}
Tian Xie, Xiang Fu, Octavian-Eugen Ganea, Regina Barzilay, and Tommi Jaakkola.
\newblock Crystal diffusion variational autoencoder for periodic material
  generation.
\newblock In \emph{International Conference on Learning Representations}, 2022.

\bibitem[Jiao et~al.(2023)Jiao, Huang, Lin, Han, Chen, Lu, and
  Liu]{jiao2023diffcsp}
Rui Jiao, Wenbing Huang, Peijia Lin, Jiaqi Han, Pin Chen, Yutong Lu, and Yang
  Liu.
\newblock Crystal structure prediction by joint equivariant diffusion.
\newblock In \emph{Advances in Neural Information Processing Systems},
  volume~36, 2023.

\bibitem[Zeni et~al.(2025)Zeni, Pinsler, Zügner, Fowler, Horton, Fu, Wang,
  Shysheya, Crabbé, Ueda, Sordillo, et~al.]{zeni2025mattergen}
Claudio Zeni, Robert Pinsler, Daniel Zügner, Andrew Fowler, Matthew Horton,
  Xiang Fu, Zilong Wang, Aliaksandra Shysheya, Jonathan Crabbé, Shoko Ueda,
  Roberto Sordillo, et~al.
\newblock A generative model for inorganic materials design.
\newblock \emph{Nature}, 639:\penalty0 624--632, 2025.
\newblock \doi{10.1038/s41586-025-08628-5}.

\bibitem[Ho and Salimans(2022)]{ho2022cfg}
Jonathan Ho and Tim Salimans.
\newblock Classifier-free diffusion guidance, 2022.

\bibitem[Xie and Grossman(2018)]{xie2018cgcnn}
Tian Xie and Jeffrey~C. Grossman.
\newblock Crystal graph convolutional neural networks for an accurate and
  interpretable prediction of material properties.
\newblock \emph{Physical Review Letters}, 120:\penalty0 145301, 2018.
\newblock \doi{10.1103/PhysRevLett.120.145301}.

\bibitem[Choudhary and DeCost(2021)]{choudhary2021alignn}
Kamal Choudhary and Brian DeCost.
\newblock Atomistic line graph neural network for improved materials property
  predictions.
\newblock \emph{npj Computational Materials}, 7:\penalty0 185, 2021.
\newblock \doi{10.1038/s41524-021-00650-1}.

\bibitem[Szymanski and Bartel(2025)]{szymanski2025baselines}
Nathan~J. Szymanski and Christopher~J. Bartel.
\newblock Establishing baselines for generative discovery of inorganic
  crystals.
\newblock \emph{Materials Horizons}, 12:\penalty0 8000--8011, 2025.
\newblock \doi{10.1039/D5MH00010F}.

\bibitem[Dobberstein and Hamaekers(2026)]{dobberstein2026materium}
Niklas Dobberstein and Jan Hamaekers.
\newblock Materium: An autoregressive approach for material generation.
\newblock \emph{Computational Materials Science}, 274:\penalty0 114980, 2026.
\newblock \doi{10.1016/j.commatsci.2026.114980}.

\bibitem[Mal et~al.(2026)Mal, Ahmed, Jami, Mishra, and Sen]{mal2026diffcrysgen}
Sourav Mal, Nehad Ahmed, Junaid Jami, Subhankar Mishra, and Prasenjit Sen.
\newblock Diffcrysgen: A generative diffusion model for accelerated design of
  inorganic crystalline materials.
\newblock \emph{npj Computational Materials}, 12:\penalty0 282, 2026.
\newblock \doi{10.1038/s41524-026-02147-1}.

\bibitem[Emery and Wolverton(2017)]{emery2017abo3}
Antoine~A. Emery and Chris Wolverton.
\newblock High-throughput dft calculations of formation energy, stability and
  oxygen vacancy formation energy of abo3 perovskites.
\newblock \emph{Scientific Data}, 4:\penalty0 170153, 2017.
\newblock \doi{10.1038/sdata.2017.153}.

\bibitem[Shannon(1976)]{shannon1976}
R.~D. Shannon.
\newblock Revised effective ionic radii and systematic studies of interatomic
  distances in halides and chalcogenides.
\newblock \emph{Acta Crystallographica Section A}, 32:\penalty0 751--767, 1976.
\newblock \doi{10.1107/S0567739476001551}.

\bibitem[Ong et~al.(2013)Ong, Richards, Jain, Hautier, Kocher, Cholia, Gunter,
  Chevrier, Persson, and Ceder]{ong2013pymatgen}
Shyue~Ping Ong, William~Davidson Richards, Anubhav Jain, Geoffroy Hautier,
  Michael Kocher, Shreyas Cholia, Dan Gunter, Vincent~L. Chevrier, Kristin~A.
  Persson, and Gerbrand Ceder.
\newblock Python materials genomics (pymatgen): A robust, open-source python
  library for materials analysis.
\newblock \emph{Computational Materials Science}, 68:\penalty0 314--319, 2013.
\newblock \doi{10.1016/j.commatsci.2012.10.028}.

\bibitem[Loshchilov and Hutter(2019)]{loshchilov2019adamw}
Ilya Loshchilov and Frank Hutter.
\newblock Decoupled weight decay regularization.
\newblock In \emph{International Conference on Learning Representations}, 2019.

\bibitem[Smith and Topin(2019)]{smith2019superconvergence}
Leslie~N. Smith and Nicholay Topin.
\newblock Super-convergence: Very fast training of neural networks using large
  learning rates.
\newblock In \emph{Artificial Intelligence and Machine Learning for
  Multi-Domain Operations Applications}, 2019.
\newblock \doi{10.1117/12.2520589}.

\bibitem[Giannozzi et~al.(2009)]{giannozzi2009qe}
Paolo Giannozzi et~al.
\newblock Quantum espresso: A modular and open-source software project for
  quantum simulations of materials.
\newblock \emph{Journal of Physics: Condensed Matter}, 21:\penalty0 395502,
  2009.
\newblock \doi{10.1088/0953-8984/21/39/395502}.

\bibitem[Giannozzi et~al.(2017)]{giannozzi2017qe}
Paolo Giannozzi et~al.
\newblock Advanced capabilities for materials modelling with quantum espresso.
\newblock \emph{Journal of Physics: Condensed Matter}, 29:\penalty0 465901,
  2017.
\newblock \doi{10.1088/1361-648X/aa8f79}.

\bibitem[Dal~Corso(2014)]{dalcorso2014pslibrary}
Andrea Dal~Corso.
\newblock Pseudopotentials periodic table: From h to pu.
\newblock \emph{Computational Materials Science}, 95:\penalty0 337--350, 2014.
\newblock \doi{10.1016/j.commatsci.2014.07.043}.

\bibitem[Togo et~al.(2024)Togo, Shinohara, and Tanaka]{togo2024spglib}
Atsushi Togo, Kohei Shinohara, and Isao Tanaka.
\newblock Spglib: A software library for crystal symmetry search.
\newblock \emph{Science and Technology of Advanced Materials: Methods},
  4:\penalty0 2384822, 2024.
\newblock \doi{10.1080/27660400.2024.2384822}.

\bibitem[Solecki and Zhu(2026)]{solecki2026bandgaps}
Evan Solecki and Peifen Zhu.
\newblock Evaluation of machine-learned band gaps: Two-stage modeling and the
  effects of data representation and split strategy.
\newblock \emph{Computational Materials Science}, 274:\penalty0 115106, 2026.
\newblock \doi{10.1016/j.commatsci.2026.115106}.

\end{thebibliography}
\end{document}